\documentclass[aps,pra,preprint,showpacs]{revtex4}
\newcommand{\PRA}[3]{Phys.~Rev.~A       {\bf #1}, #2 (#3).}

\begin{document}
\preprint{Version 1.2}

\title{Singular Hylleraas three-electron integrals}

\author{Krzysztof Pachucki}
\email[]{krp@fuw.edu.pl}
\affiliation{Institute of Theoretical Physics,
             University of Warsaw,
             Ho\.{z}a 69, 00-681 Warsaw, Poland}

\author{Mariusz Puchalski}
\email[]{mpuchals@fuw.edu.pl}
\affiliation{Physics Department,
             University of Alberta,
             11322 89 Avenue, Edmonton, AB, T6G 2G7 Canada}


\begin{abstract}
Calculations of the leading quantum electrodynamics effects in
few electron systems involve singular matrix elements of
the inter-electronic distances of the form $1/r_i^3$ and $1/r_{ij}^3$.
Integrals that result when the nonrelativistic wave
function is represented with a Hylleraas basis
representation are studied.
Recursion relations for various powers
of the electron coordinates and the master integrals are derived
in a form suited for high precision numerical evaluations.
\end{abstract}

\maketitle
\section{Introduction}
A challenging task in high precision calculations of energy levels
of few electron atoms or ions is the  accurate solution $\psi$ of
the nonrelativistic Schr\"odinger equation.
This wave function $\psi$ is next used to obtain relativistic and
quantum electrodynamics (QED) effects including finite nuclear mass corrections.
The most accurate representation of the three-electron
wave function $\psi$ achieved so far \cite{lit_wave1, lit_wave2}
uses the Hylleraas basis set \cite{hyll}, namely
\begin{eqnarray}
\psi &=& \sum_i c_i\,\phi_i\nonumber \\
\phi &=&
e^{-w_1\,r_1-w_2\,r_2-w_3\,r_3}\,r_{23}^{n_1}\,r_{31}^{n_2}\,r_{12}^{n_3}\,
r_1^{n_4}\,r_2^{n_5}\,r_3^{n_6},
\label{00}
\end{eqnarray}
with $n_i$ being nonnegative integers. This basis set allows the Kato
cusp condition to be satisfied to a high degree, thus ensuring good
convergence for matrix elements of relativistic and QED operators.
All these matrix elements can be expressed in terms of the
Hylleraas integrals $f$, see Eq. (\ref{01}). The calculation of
the nonrelativistic energy involves integrals with nonnegative
$n_i$. Such integrals have been worked out by King {\em et al.} in
the series of works \cite{king_lit_nrel}, and more recently by Yan and
Drake \cite{lit_wave1, yandrake_lit_nrel}, Sims and Hegstrom \cite{sims}, and by us in
Ref. \cite{recursions}. Our approach relies on analytic recursion
relations, which are highly efficient and
sufficiently stable to achieve accurate numerical results.

Matrix elements of relativistic operators involve the so called
extended Hylleraas integral, where one of $n_i=-1$ (see Eq. (\ref{01})). These
integrals have been worked out first by King {\em et al.} in
\cite{king_lit_sing}, and later by Yan and Drake in Refs. \cite{yandrake_lit_sing}
and by us in \cite{rec_sing, lit_wave2}. The use of recursion
relations allows for a significant increase in the size of the
basis set and the accuracy of the obtained results
\cite{lit_wave2}. Finally, matrix elements of QED operators involve
integrals with $n_i=-2$. Such integrals have been worked out by
Yan in \cite{yan}, but no numerical results for any particular
integral have been published, which would serve as a test of achieved accuracy.
In this work we develop recursion
relations for three-electron Hylleraas integrals either with
$1/r_{ij}^3$ or with $1/r_i^3$, and obtain a one dimensional integral
representation for the master integral which is suitable for 
precise numerical evaluation.

In Section II we recall known results for the regular Hylleraas
integrals, in Section III we treat integrals involving $1/r_i^2$,
and in Section IV we treat integrals involving $1/r_{ij}^2$. Apart from known results,
we present also a new form of the master integral with the hope
that it can be used in the future to obtain a fully analytic result.
In Section V we develop recursion relations for
integrals involving $1/r_i^3$ which are very similar to those
involving $1/r_i^2$. In Section VI, which is the most difficult one,
we obtain a full set of recursions for Hylleraas integrals with $1/r_{ij}^3$, and
in Section VII we present numerical results together with a short summary.

\section{Regular three-electron Hylleraas integral}
The regular three-electron Hylleraas integral is
\begin{eqnarray}
f(n_1,n_2,n_3;n_4, n_5, n_6) &=& \int\frac{d^3\,r_1}{4\,\pi}\int\frac{d^3\,r_2}{4\,\pi}
\int\frac{d^3\,r_3}{4\,\pi}\,e^{-w_1\,r_1-w_2\,r_2-w_3\,r_3}\nonumber \\ &&
r_{23}^{n_1-1}\,r_{31}^{n_2-1}\,r_{12}^{n_3-1}\,
r_{1}^{n_4-1}\,r_{2}^{n_5-1}\,r_{3}^{n_6-1}\,, \label{01}
\end{eqnarray}
with all $n_i \geq 0$. The most convenient way to perform this integral
is by using recursion relations. The initial values with $n_i=0,1$
are known explicitly \cite{remiddi, recursions}.
\begin{eqnarray}
f(0,0,0;0,0,0) &=& \frac{-1}{2\,w_1\,w_2\,w_3}\biggl[
{\rm Lg}\biggl(\frac{w_3}{w_1+w_2}\biggr)+
{\rm Lg}\biggl(\frac{w_2}{w_3+w_1}\biggr)+
{\rm Lg}\biggl(\frac{w_1}{w_2+w_3}\biggr)\biggr]\,,\label{02}\\
f(1,0,0;0,0,0) &=& -\frac{1}{w_2^2\,w_3^2}\,\ln\biggl[
                    \frac{w_1\,(w_1+w_2+w_3)}{(w_1+w_2)(w_1+w_3)}\biggr]\,,\label{03}\\
f(0,1,0;0,0,0) &=& -\frac{1}{w_1^2\,w_3^2}\,\ln\biggl[
                    \frac{w_2\,(w_1+w_2+w_3)}{(w_2+w_3)(w_2+w_1)}\biggr]\,,\label{04}\\
f(0,0,1;0,0,0) &=& -\frac{1}{w_1^2\,w_2^2}\,\ln\biggl[
                    \frac{w_3\,(w_1+w_2+w_3)}{(w_3+w_1)(w_3+w_2)}\biggr]\,,\label{05}\\
f(1,1,0;0,0,0) &=& \frac{1}{w_1\,w_2\,(w_1+w_2)\,w_3^2}\,,\label{06}\\
f(1,0,1;0,0,0) &=& \frac{1}{w_1\,w_3\,(w_1+w_3)\,w_2^2}\,,\label{07}\\
f(0,1,1;0,0,0) &=& \frac{1}{w_2\,w_3\,(w_2+w_3)\,w_1^2}\,,\label{08}\\
f(1,1,1;0,0,0) &=& \frac{1}{w_1^2\,w_2^2\,w_3^2}\,,\label{09}
\end{eqnarray}
where
\begin{equation}
{\rm Lg}(x) = {\rm Li}_2(1-x) + {\rm Li}_2(-x)+\ln(x)\,\ln(1+x)\,,\label{10}
\end{equation}
and Li$_2(x)$ is the dilogarithmic function. All other Hylleraas
integrals for arbitrary but nonnegative $n_i$ can be obtained by
6 independent recursion relations for each $n_i$, see Ref.
\cite{recursions}. The numerical stability of these recursions is
an issue. For the maximal number $\Omega=\sum_i n_i = 30$ we used
sextuple precision arithmetic and verified against octuple
precision that results for nonrelativistic energies are
numerically significant for at least 16 digits.

\section{Extended Hyleraas integral with $1/r_i^2$}
More difficult to evaluate are extended Hylleraas
three-electron integrals involving additional single powers of
$r_i$ or $r_{ij}$ in the denominator. The first kind of integral
with  $1/r_i^2$ can be obtained by integration with respect to
a corresponding parameter $w_i$, namely
\begin{equation}
f(0,0,0;-1,n_5,n_6) = \int_{w_1}^\infty dw_1\, f(0,0,0;0,n_5,n_6),\label{11}
\end{equation}
where $f(0,0,0;0,n_5,n_6)$ is obtained by recursions which are
numerically stable at high values of $w_1$. The integration in Eq.
(\ref{11}) is performed using the adapted quadrature \cite{rokhlin} for the class
of functions involving logarithms \cite{rokhlin}, which allows one
to achieve high precision at low evaluation cost. The function
$f(n_1,n_2,n_3;-1,n_5,n_6)$ for arbitrary integer values of $n_1,
n_2$ and $n_3$ can be obtained \cite{lit_wave2} by recursion
relations starting from $f(0,0,0;-1,n_5,n_6)$.

\section{Extended Hyleraas integral with $1/r_{ij}^2$}
The second kind of extended Hylleraas integral involves $1/r_{ij}^2$, and
we will investigate it here in more detail. Recursion relations
start from the master integral
\begin{eqnarray}
h(w_1,w_2,w_3) &\equiv& f(-1,0,0;0,0,0)\nonumber \\
&=& \int\frac{d^3\,r_1}{4\,\pi}\int\frac{d^3\,r_2}{4\,\pi}
\int\frac{d^3\,r_3}{4\,\pi}\,e^{-w_1\,r_1-w_2\,r_2-w_3\,r_3}\,
r_{23}^{-2}\,r_{31}^{-1}\,r_{12}^{-1}\,
r_{1}^{-1}\,r_{2}^{-1}\,r_{3}^{-1}\,.\label{12}
\end{eqnarray}
and some other simpler two electron-like integrals. 
By using integration by parts identities \cite{parts},
the following differential equations for the {\em h} function 
have been obtained in Ref. \cite{rec_sing}
\begin{eqnarray}
\frac{\partial h}{\partial w_2} &=& \frac{-1}{2\,w_1^2\,w_2}
\Bigl[F + (2\,w_1^2 + w_2^2 - w_3^2)\,h
+ w_1\,(w_1^2 + w_2^2 - w_3^2)\,\frac{\partial h}{\partial w_1}\Bigr] ,\label{13}\\
\frac{\partial h}{\partial w_3} &=& \frac{-1}{2\,w_1^2\,w_3}
\Bigl[-F + (2\,w_1^2 - w_2^2 + w_3^2)\,h
+ w_1\,(w_1^2 - w_2^2 + w_3^2)\,\frac{\partial h}{\partial w_1}\Bigr] ,\label{14}
\end{eqnarray}
and
\begin{equation}
w\,w_1^2\,\frac{\partial^2 h}{\partial w_1^2} +
w_1\,\bigl[4\,w_1^2\,(w_1^2 - w_2^2 - w_3^2)+w\bigr]\,\frac{\partial h}{\partial w_1}
+\bigl[w_1^4 + 4\,w_1^2\,(w_1^2 - w_2^2 - w_3^2)-w\bigr]\,h(w_1) = R,\label{15}
\end{equation}
where
\begin{eqnarray}
R &=& w_1\,w_2\,\ln\biggl(1 + \frac{w_1}{w_2}\biggr) +
           w_1\,w_3\,\ln\biggl(1 + \frac{w_1}{w_3}\biggr) +
           (w_2^2 - w_3^2)\,\ln\biggl(\frac{w_1 + w_3}{w_1 + w_2}\biggr)
\nonumber \\ &&
           +2\,w_1^2\,\ln\biggl(\frac{w_1\,(w_1 + w_2 + w_3)}{(w_1 + w_2)\,(w_1 + w_3)}\biggr) +
           (w_2^2 - w_3^2)\,F ,\label{16}\\
F &=&\frac{1}{2}\biggl[
2\,{\rm Li}_2\biggl( -\frac{w_2}{w_1}\biggr) - {\rm Li}_2\biggl(1 - \frac{w_2}{w_3}\biggr) +
    {\rm Li}_2\biggl(1 - \frac{w_1 + w_2}{w_3}\biggr) - 2\,{\rm Li}_2\biggl(-\frac{w_3}{w_1}\biggr)
\nonumber \\ &&
    + {\rm Li}_2\biggl(\frac{w_2}{w_2 + w_3}\biggr) - {\rm Li}_2\biggl(\frac{w_3}{w_2 + w_3}\biggr) +
    {\rm Li}_2\biggl(\frac{w_2}{w_1 + w_2 + w_3}\biggr) -
    {\rm Li}_2\biggl(\frac{w_3}{w_1 + w_2 + w_3}\biggr)
\nonumber \\ &&
    + {\rm Li}_2\biggl(1 - \frac{w_3}{w_2}\biggr) - {\rm Li}_2\biggl(1 -
    \frac{w_1 + w_3}{w_2}\biggr)
+\ln\biggl(\frac{w_2}{w_3}\biggr)\,\ln\biggl(\frac{w_1 + w_2 + w_3}{w_2 +
  w_3}\biggr)\biggr],\label{17}\\
w &=& (w_1 - w_2 - w_3) (w_1 - w_2 + w_3) (w_1 + w_2 - w_3)  (w_1 + w_2 + w_3).\label{18}
\end{eqnarray}
It is convenient to express the function $h$ 
\begin{equation}
h(w_1,w_2,w_3) = \frac{1}{w_1\,\sqrt{w_2\,w_3}}\,
H\biggl(\frac{w_1^2-w_2^2-w_3^2}{4\,w_2\,w_3},\frac{w_2}{w_3}\biggr),\label{19}
\end{equation}
in terms of a dimensionless function $H(x,y)$ of two variables $x$ and $y$.
Then the first two differential equations become apparently
equivalent and take the form
\begin{equation}
\frac{\partial H}{\partial y} =
-\frac{1}{2\,y}\,\biggl(4\,x +y+\frac{1}{y}\biggr)^{-1/2}\,F,\label{20}
\end{equation}
where
\begin{equation}
F = F(w_1,w_2,w_3) = F\biggl(\sqrt{4\,x+y+\frac{1}{y}},\sqrt{y},\frac{1}{\sqrt{y}}\biggr),
\label{21}
\end{equation}
and the third differential equation, with respect to $w_1$ becomes
\begin{equation}
(4\,x^2-1)\,\frac{\partial^2 H}{\partial x^2} +
8\,x\,\frac{\partial H}{\partial x} + H = R'(x,y),\label{22}
\end{equation}
where
\begin{eqnarray}
R'(x,y) &=& \frac{\sqrt{w_2\,w_3}}{w_1^3}\,R(w_1,w_2,w_3)\nonumber \\
              &=& \biggl(4\,x+y+\frac{1}{y}\biggr)^{-3/2}\,
                  R\biggl(\sqrt{4\,x+y+\frac{1}{y}},\sqrt{y},\frac{1}{\sqrt{y}}\biggr).
\label{23}
\end{eqnarray}
The homogenous differential equation in (\ref{22}) is satisfied by
complete elliptic integrals $K(1/2\pm x)$, and
the solution of the inhomogeneous equation is obtained by Euler's method of
variation of constants \cite{rec_sing}
\begin{eqnarray}
H(x,y) &=& \frac{1}{\pi}\,\biggl[
K\biggl(\frac{1}{2}+x\biggr)\,\int_x^{1/2} dz\,R'(z,y)
\,K\biggl(\frac{1}{2}-z\biggr) \nonumber \\ &&
+K\biggl(\frac{1}{2}-x\biggr)\,\int_{-1/2}^x dz\,R'(z,y)
\,K\biggl(\frac{1}{2}+z\biggr)\biggr].\label{24}
\end{eqnarray}
This integral form is convenient for the numerical calculation of
the master integral $h(w_1,w_2,w_3)$ and its derivatives.
However, a different and apparently simpler form can be obtained from Eq. (\ref{20})
\begin{eqnarray}
H\biggl(x,\frac{w_2}{w_3}\biggr) &=& \int_0^{w_2/w_3} dy \,\biggl(-\frac{1}{2\,y}\biggr)\,
\biggl(4\,x+y+\frac{1}{y}\biggr)^{-1/2}\,F\nonumber \\
       &=&
 \int_0^{\sqrt{w_2/w_3}}
 \frac{du}{u}\,\frac{1}{\sqrt{4\,x'+\bigl(u-1/u\bigr)^2}}\,
 F\bigl(\sqrt{4\,x'+\bigl(u-1/u)^2},1/u,u\bigr),\label{25}
\end{eqnarray}
where $x'=x+1/2$. This integral can be further transformed
to a form involving Jacobi elliptic functions,
but we have not been able to perform it analytically.
In particular cases, where $x=\pm 1/2$, this integral becomes ($w_2<w_3$)
\begin{eqnarray}
H\biggl(\frac{1}{2},\frac{w_2}{w_3}\biggr) &=& \int_0^{\sqrt{w_2/w_3}} du
\,\frac{1}{1+u^2}\, F\bigl(1+u^2,1,u^2\bigr),\label{26}\\
H\biggl(-\frac{1}{2},\frac{w_2}{w_3}\biggr) &=&\int_0^{\sqrt{w_2/w_3}} du
\,\frac{1}{1-u^2}\, F\bigl(1-u^2,1,u^2\bigr),\label{27}
\end{eqnarray}
and this can be expressed in terms of ${\rm Li}_3$, ${\rm Li}_2$, and
logarithmic functions. We use Eqs. (\ref{26},\ref{27}) in cases 
$w_1>w_2+w_3$ and $w_1<|w_2-w_3|$, and
for its numerical calculations we employ adapted quadrature \cite{rokhlin}
with 120 points, to achieve about  64 significant digits accuracy.

Recursion relations for $f(-1,n_2,n_3,n_4,n_5,n_6)$ have been derived in \cite{rec_sing}.
Among them, the most numerically unstable is the one which increases
the parameter $n_4$,  see Eq. (36) of Ref. \cite{rec_sing}. Namely, 
close to singularity points, $x=\pm 1/2$ and for small $\omega_1$,
we use Taylor expansion to avoid numerical instabilities.

\section{Singular Hylleraas integral with $1/r_i^3$}
The singular Hylleraas integrals which involve $1/r_i^3$ and $1/r_{ij}^3$
are needed for the computation of QED effects \cite{yan}.
Let us first define a distribution $P(1/r^3)$
\begin{equation}
\langle\phi| P\left(\frac{1}{r^3}\right)|\psi\rangle =
\lim_{\epsilon\rightarrow 0}\int {\rm d}^3 r\,
\phi^*(\vec r)\left[\frac{1}{r^3}\,\Theta(r-\epsilon)
+ 4\,\pi\,\delta^3(r)\,
(\gamma+\ln \epsilon)\right]\,\psi(\vec r)\,. \label{28}
\end{equation}
Any factor $1/r^3$ in the following will be understood 
to be defined in the above sense,
and we will drop the symbol $ P()$.
It follows from this definition that
\begin{equation}
\biggl\langle\frac{e^{-w\,r}}{r^3}\biggr\rangle = C\,\ln(w) + O(1/\omega)\label{29}
\end{equation}
for large $\omega$. Moreover, we will use in the derivation below the
following integral representations  
\begin{eqnarray}
\frac{1}{r^3} &=&
\lim_{\Lambda\rightarrow\infty}\biggl\{\int_0^\Lambda dt\,t\,\frac{e^{-t\,r}}{r}
+4\,\pi\,\delta^3(r)\,(1-\ln\Lambda)\biggr\}\nonumber \\ &=&
\frac{1}{2}\,\lim_{\Lambda\rightarrow\infty}\biggl\{\int_0^\Lambda dt\,t^2\,e^{-t\,r}
+4\,\pi\,\delta^3(r)\,[2\,(1-\ln\Lambda)+1]\biggr\}.\label{30}
\end{eqnarray}

To obtain recursion relations for the singular three-electron Hylleraas
integral one first considers the integral $G$
\begin{eqnarray}
G(m_1,m_2,m_3;m_4,m_5,m_6) &=& \frac{1}{8\,\pi^6}\,\int d^3k_1\int d^3k_2\int d^3k_3\,
(k_1^2+u_1^2)^{-m_1}\,(k_2^2+u_2^2)^{-m_2} \nonumber\\ &&\hspace*{-1cm}(k_3^2+u_3^2)^{-m_3}\,
(k_{32}^2+w_1^2)^{-m_4}\,(k_{13}^2+w_2^2)^{-m_5}\,(k_{21}^2+w_3^2)^{-m_6}, \label{32}
\end{eqnarray}
which is related to $f$ by $f(0,0,0,0,0,0) = G(1,1,1,1,1,1)|_{u_1=u_2=u_3=0}$.
The following 9 integration by parts (IBP) identities are valid
because the integral of the derivative of a function
vanishing at infinity vanishes,
\begin{eqnarray}
&&0 \equiv {\rm id}(i,j) =
\int d^3k_1\int d^3k_2\int d^3k_3\,\frac{\partial}{\partial\,{\vec k_i}}
 \Bigl[ \vec k_j\,(k_1^2+u_1^2)^{-1}
\nonumber \\ &&
(k_2^2+u_2^2)^{-1}\,(k_3^2+u_3^2)^{-1}
(k_{32}^2+w_1^2)^{-1}\,(k_{13}^2+w_2^2)^{-1}\,(k_{21}^2+w_3^2)^{-1}
\Bigr] ,
\label{33}
\end{eqnarray}
where $i,j=1,2,3$.
The reduction of the scalar products from the numerator leads to 
identities for linear combination of the $G$ functions.
If any of the arguments is equal to 0, then $G$ becomes a known two-electron
Hylleraas type integral. These identities can be used to derive various
recursion relations. Here we derive a set of recursions for $f(n_1,n_2,n_3,-2,n_5,n_6)$.
This is achieved in a few steps. In the first step we use
integration by parts identities in momentum representation Eq. (\ref{33}),
to form the following linear combination
\begin{eqnarray}
{\rm id}(2,2)+{\rm id}(3,3)-{\rm id}(1,1) &=& 2\,\bigl[
G(0, 1, 1, 1, 1, 2) + G(0, 1, 1, 1, 2, 1) - G(1, 0, 1, 1, 1, 2)
\nonumber \\ &&
- G(1, 1, 0, 1, 2, 1) - G(1, 1, 1, 1, 1, 1)/2 - G(2, 1, 1, 1, 1, 1)\,u_1^2
\nonumber \\ &&
-G(1, 1, 1, 1, 1, 2)\,(u_1^2 - u_2^2) + G(1, 2, 1, 1, 1, 1)\,u_2^2
\nonumber \\ &&
- G(1, 1, 1, 1, 2, 1)\,(u_1^2 - u_3^2) + G(1, 1, 2, 1, 1, 1)\,u_3^2
\nonumber \\ &&
+ G(1, 1, 1, 2, 1, 1)\,w_1^2\bigr] = 0\,.\label{34}
\end{eqnarray}
We use Eq. (\ref{30}) to integrate with respect to $w_1$ and differentiate over $u_1$, $u_2$, $u_3$,
$w_2$, and $w_3$ to obtain the main formula
\begin{eqnarray}
f(n_1, n_2, n_3, -2, n_5, n_6) &=& \frac{1}{(n_2+n_3-n_1-1)\,w_2\,w_3}\,\Bigl[
\nonumber \\ &&
(n_1-1)\,n_1\,n_5\,f(n_1-2, n_2, n_3, -2, n_5-1, n_6+1)
\nonumber \\ &&
+ (n_1-1)\,n_1\,n_6\,f(n_1-2, n_2, n_3, -2, n_5+1, n_6-1)
\nonumber \\ &&
- (n_2-1)\,n_2\,n_5\,f(n_1, n_2-2, n_3, -2, n_5-1, n_6+1)
\nonumber \\ &&
- (n_3-1)\,n_3\,n_6\,f(n_1, n_2, n_3-2, -2, n_5+1, n_6-1)
\nonumber \\ &&
+ (n_1 - n_2 - n_3+1)\,n_5\,n_6\,f(n_1, n_2, n_3, -2, n_5-1, n_6-1)
\nonumber \\ &&
+ n_5\,n_6\,f(n_1, n_2, n_3, -1, n_5-1, n_6-1)\,w_1
\nonumber \\ &&
- (n_1-1)\,n_1\,f(n_1-2, n_2, n_3, -2, n_5, n_6+1)\,w_2
\nonumber \\ &&
+ (n_2-1)\,n_2\,f(n_1, n_2-2, n_3, -2, n_5, n_6+1)\,w_2
\nonumber \\ &&
- (n_1 - n_2 - n_3+1)\,n_6\,f(n_1, n_2, n_3, -2, n_5, n_6-1)\,w_2
\nonumber \\ &&
- n_6\,f(n_1, n_2, n_3, -1, n_5, n_6-1)\,w_1\,w_2
\nonumber \\ &&
- (n_1-1)\,n_1\,f(n_1-2, n_2, n_3, -2, n_5+1, n_6)\,w_3
\nonumber \\ &&
+ (n_3-1)\,n_3\,f(n_1, n_2, n_3-2, -2, n_5+1, n_6)\,w_3
\nonumber \\ &&
- (n_1 - n_2 - n_3+1)\,n_5\,f(n_1, n_2, n_3, -2, n_5-1, n_6)\,w_3
\nonumber \\ &&
- n_5\,f(n_1, n_2, n_3, -1, n_5-1, n_6)\,w_1\,w_3
\nonumber \\ &&
+ f(n_1, n_2, n_3, -1, n_5, n_6)\,w_1\,w_2\,w_3
\nonumber \\ &&
- n_5\,\delta(n_1)\,\Gamma(n_5 + n_6-1, -2, n_2 + n_3-1, w_2 + w_3, w_1, 0)
\nonumber \\ &&
- n_6\,\delta(n_1)\,\Gamma(n_5 + n_6-1, -2, n_2 + n_3-1, w_2 + w_3, w_1, 0)
\nonumber \\ &&
+ \delta(n_1)\,\Gamma(n_5 + n_6, -2, n_2 + n_3-1, w_2 + w_3, w_1, 0)\,w_2
\nonumber \\ &&
+ \delta(n_1)\,\Gamma(n_5 + n_6, -2, n_2 + n_3-1, w_2 + w_3, w_1, 0)\,w_3
\nonumber \\ &&
+ n_5\,\delta(n_2)\,\Gamma(n_6-2, n_5-1, n_1 + n_3-1, w_1 + w_3, w_2, 0)
\nonumber \\ &&
- \delta(n_2)\,\Gamma(n_6-2, n_5, n_1 + n_3-1, w_1 + w_3, w_2, 0)\,w_2
\nonumber \\ &&
+ n_6\,\delta(n_3)\,\Gamma(n_5-2, n_6-1, n_1 + n_2-1, w_1 + w_2, w_3, 0)
\nonumber \\ &&
- \delta(n_3)\,\Gamma(n_5-2, n_6, n_1 + n_2-1, w_1 + w_2, w_3, 0)\,w_3
\nonumber \\ &&
- n_5\,n_6\,\Gamma(n_3 + n_5-2, n_2 + n_6-2, n_1, w_2, w_3, 0)
\nonumber \\ &&
+ n_6\,\Gamma(n_3 + n_5-1, n_2 + n_6-2, n_1, w_2, w_3, 0)\,w_2
\nonumber \\ &&
+ n_5\,\Gamma(n_3 + n_5-2, n_2 + n_6-1, n_1, w_2, w_3, 0)\,w_3
\nonumber \\ &&
- \Gamma(n_3 + n_5-1, n_2 + n_6-1, n_1, w_2, w_3, 0)\,w_2\,w_3
\Bigr]\,.\label{35}
\end{eqnarray}
It takes a particularly simple form when all $n_i$ are equal to $0$:
\begin{eqnarray}
f(0, 0, 0, -2, 0, 0) &=& 
\frac{1}{w_2\,w_3}\,\Bigl[
- w_1\,w_2\,w_3\,f(0, 0, 0, -1, 0, 0) 
+ w_3\,\Gamma(-2, 0, -1, w_1 + w_2, w_3, 0) \nonumber \\ &&
+ w_2\,\Gamma(-2, 0, -1, w_1 + w_3, w_2, 0) 
+ w_2\,w_3\,\Gamma(-1, -1, 0, w_2, w_3, 0) \nonumber \\ &&
- (w_2+w_3)\,\Gamma(0, -2, -1, w_2 + w_3, w_1, 0)\Bigr] \nonumber \\ &=&
-w_1\,f(0, 0, 0, -1, 0, 0) 
- w_1\,(w_1+w_2+w_3)\,f(0, 0, 0, 0, 0, 0)  \label{36}\\ &&\hspace*{-18ex}
+\frac{1}{w_2}\biggl[\bigl(2 - \ln(w_1)\bigr)\,
\ln\biggl(1 + \frac{w_2}{w_3}\biggr) +
\frac{1}{2}\,\ln^2\biggl(\frac{w_3}{w_1}\biggr) +
{\rm Li}_2\biggl(1 - \frac{w_1 + w_2}{w_3}\biggr)
+ {\rm Li}_2\biggl(1 - \frac{w_2 + w_3}{w_1}\biggr)\biggr]\nonumber \\ &&
\hspace*{-18ex}+\frac{1}{w_3}\biggl[\bigl(2 - \ln(w_1)\bigr)\,
\ln\biggl(1 + \frac{w_3}{w_2}\biggr) +
\frac{1}{2}\,\ln^2\biggl(\frac{w_2}{w_1}\biggr) +
{\rm Li}_2\biggl(1 - \frac{w_1 + w_3}{w_2}\biggr)
+ {\rm Li}_2\biggl(1 - \frac{w_2 + w_3}{w_1}\biggr)\biggr],
\nonumber
\end{eqnarray}
where $\Gamma$ is the two electron Hylleraas integral, defined in the appendix.
The general formula in Eq. (\ref{35}) does not work in the case $1+n_1=n_2+n_3$.
In this special case we use IBP identities in coordinate
space  and limit ourselves only to identities of the form
\begin{equation}
0 \equiv {\rm id}(i) = \int d^3 r_1\,\int d^3 r_2\,\int d^3 r_3\,
\bigl(g\,\nabla^2_i h - h\,\nabla^2_i g\bigr)\,,
\label{37}
\end{equation}
where
\begin{eqnarray}
g &=& e^{-w_1\,r_1-w_2\,r_2-w_3\,r_3}\,r_1^{n_4-1}\,r_2^{n_5-1}\,r_3^{n_6-1}\,,
\label{38} \\
h &=& r_{23}^{n_1-1}\,r_{31}^{n_2-1}\,r_{12}^{n_3-1}\,.
\label{39}
\end{eqnarray}
The identities id$(2)$ and id$(3)$
\begin{eqnarray}
f(n_1, n_2, n_3, -2, n_5, n_6) &=&\frac{1}{w_2^2}\,
\Bigl[(n_1-1)\,(n_1 + n_3-1)\,f(n_1-2, n_2, n_3, -2, n_5, n_6)
\nonumber \\ &&
- (n_1-1)\,(n_3-1)\,f(n_1-2, 2 + n_2, n_3-2, -2, n_5, n_6)
\nonumber \\ &&
+(n_3-1)\,(n_1 + n_3-1)\,f(n_1, n_2, n_3-2, -2, n_5, n_6)
\nonumber \\ &&
- (n_5-1)\,n_5\,f(n_1, n_2, n_3, -2, n_5-2, n_6)
\nonumber \\ &&
+ 2\,n_5\,f(n_1, n_2, n_3, -2, n_5-1, n_6)\,w_2
\nonumber \\ &&
+\delta(n_5)\,\Gamma(n_1 + n_6-1, n_3 -3, n_2, w_3, w_1, 0)\Bigr],
\label{40} \\
f(n_1, n_2, n_3, -2, n_5, n_6) &=&\frac{1}{w_3^2}\,
\Bigl[-(n_1-1)\,(n_2-1)\,f(n_1-2, n_2-2, n_3+2, -2, n_5, n_6)
\nonumber \\ &&
+(n_1-1)\,(n_1 + n_2-1)\,f(n_1-2, n_2, n_3, -2, n_5, n_6)
\nonumber \\ &&
+ (n_2-1)\,(n_1 + n_2-1)\,f(n_1, n_2-2, n_3, -2, n_5, n_6)
\nonumber \\ &&
-(n_6-1)\,n_6\,f(n_1, n_2, n_3, -2, n_5, n_6-2)
\nonumber \\ &&
+ 2\,n_6\,f(n_1, n_2, n_3, -2, n_5, n_6-1)\,w_3
\nonumber \\ &&
+ \delta(n_6)\,\Gamma(n_2 -3, n_1 + n_5-1, n_3, w_1, w_2, 0)
\Bigr], \label{41}
\end{eqnarray}
replace the main recursion in Eq. (\ref{35}) for the case $1+n_1=n_2+n_3$, and can be
used also for all other $n_i$ under the conditions that $n_1>0$, $n_3>0$ or
$n_1>0$, $n_2>0$, respectively.
The case of $n_1=0, n_2=0,n_3=1$ or $n_1=0, n_2=1,n_3=0$,
which is not covered by above recursions, remains to be treated. Therefore,
in the third step one obtains the necessary recursions from
the following combination of IBP identities in the momentum space:
id$(1,1)$ +  id$(2,1)$ + id$(3,1) \equiv 0$
\begin{eqnarray}
f(0, 1, 0, -2, n_5, n_6) &=& -\frac{1}{w_3^2}\,\Bigl[
(n_6-1)\,n_6\,f(0, 1, 0, -2, n_5, n_6-2)
\nonumber \\ &&
- 2\,n_6\,f(0, 1, 0, -2, n_5, n_6-1)\,w_3
\nonumber \\ &&
+ \Gamma(n_5 + n_6-1, -2, 0, w_2 + w_3, w_1, 0)
\nonumber \\ &&
- \delta(n_6)\,\Gamma(-2, n_5-1, 0, w_1, w_2, 0)
\Bigr],\label{42}\\
f(0, 0, 1, -2, n_5, n_6) &=& -\frac{1}{w_2^2}\,\Bigl[
(n_5-1)\,n_5\,f(0, 0, 1, -2, n_5-2, n_6)
\nonumber \\ &&
- 2\,n_5\,f(0, 0, 1, -2, n_5-1, n_6)\,w_2
\nonumber \\ &&
+ \Gamma(n_5 + n_6-1, -2, 0, w_2 + w_3, w_1, 0)
\nonumber \\ &&
- \delta(n_5)\,\Gamma(n_6-1, -2, 0, w_3, w_1, 0)
\Bigr].\label{43}
\end{eqnarray}
This completes the evaluation of singular Hylleraas integrals involving $1/r_i^3$.

\section{Singular Hylleraas integral with $1/r_{ij}^3$}
The derivation of recursion relations and the master integral is similar
to the one for Hylleraas integral with $1/r_{ij}^2$, see Ref. \cite{rec_sing}.
In the first step of deriving recursion relations
we take the difference ${\rm id}(3,2)-{\rm id}(2,2)$
and use it as an equation for $G(1, 2, 1, 1, 1, 1)$,
\begin{eqnarray}
G(1, 2, 1, 1, 1, 1)\,(u_2^2 - u_3^2 + w_1^2) &=&
                         G(1, 1, 1, 0, 1, 2) - G(1, 1, 1, 1, 0, 2)
                \nonumber \\ &&
                        + G(1, 1, 1, 1, 1, 1) - G(1, 2, 0, 1, 1, 1)
                \nonumber \\ &&
                        + G(1, 2, 1, 0, 1, 1) - 2\,G(1, 1, 1, 2, 1, 1)\,w_1^2
                \nonumber \\ &&
                        + G(1, 1, 1, 1, 1, 2)\,(w_2^2 - w_1^2 - w_3^2) .
\label{44}
\end{eqnarray}
Similarly, the difference ${\rm id}(2,3)-{\rm id}(3,3)$ is used to obtain
$G(1, 1, 2, 1, 1, 1)$.
These two equations are used now to derive recursions in $n_2$ and $n_3$.
With the help of Eq. (\ref{30})
one integrates Eq. (\ref{44}) with respect to $u_1$, which lowers the first argument $n_1$ to $-2$.
Next, one differentiates with respect to
$u_2,u_3,w_1,w_2,w_3$ at $u_2=u_3=0$ to generate arbitrary powers
of $r_{13}, r_{12}, r_1, r_2, r_3$ and obtains the recursion relation in $n_2$
\begin{eqnarray}
f(-2, n_2+1, n_3, n_4, n_5, n_6) &=& \frac{1}{w_1^2\,w_3}\,\Bigl[
n_2\,(n_4-1)\,n_4\,f(-2, n_2-1, n_3, n_4-2, n_5, n_6+1)
\nonumber \\ &&
   -2\,n_2\,n_4\,w_1\,f(-2, n_2-1, n_3, n_4-1, n_5, n_6+1)
\nonumber \\ &&
   -n_2\,(n_5-1)\,n_5\,f(-2, n_2-1, n_3, n_4, n_5-2, n_6+1)
\nonumber \\ &&
   +2\,n_2\,n_5\,w_2\,f(-2, n_2-1, n_3, n_4, n_5-1, n_6+1)
\nonumber \\ &&
   +n_2\,n_6\,(n_2 + 2\,n_4 + n_6)\,f(-2, n_2-1, n_3, n_4, n_5, n_6-1)
\nonumber \\ &&
   -n_2\,(n_2 + 2\,n_4 + 2\,n_6+1)\,w_3\,f(-2, n_2-1, n_3, n_4, n_5, n_6)
\nonumber \\ &&
   +n_2\,(w_1^2 - w_2^2 + w_3^2)\,f(-2, n_2-1, n_3, n_4, n_5, n_6+1)
\nonumber \\ &&
   -2\,n_2\,n_6\,w_1\,f(-2, n_2-1, n_3, n_4+1, n_5, n_6-1)
\nonumber \\ &&
   +2\,n_2\,w_1\,w_3\,f(-2, n_2-1, n_3, n_4+1, n_5, n_6)
\nonumber \\ &&
   -(n_3-1)\,n_3\,n_6\,f(-2, n_2+1, n_3-2, n_4, n_5, n_6-1)
\nonumber \\ &&
   +(n_3-1)\,n_3\,w_3\,f(-2, n_2+1, n_3-2, n_4, n_5, n_6)
\nonumber \\ &&
   +(n_4-1)\,n_4\,n_6\,f(-2, n_2+1, n_3, n_4-2, n_5, n_6-1)
\nonumber \\ &&
   -(n_4-1)\,n_4\,w_3\,f(-2, n_2+1, n_3, n_4-2, n_5, n_6)
\nonumber \\ &&
   -2\,n_4\,n_6\,w_1\,f(-2, n_2+1, n_3, n_4-1, n_5, n_6-1)
\nonumber \\ &&
   +2\,n_4\,w_1\,w_3\,f(-2, n_2+1, n_3, n_4-1, n_5, n_6)
\nonumber \\ &&
   +n_6\,w_1^2\,f(-2, n_2+1, n_3, n_4, n_5, n_6-1)
\nonumber \\ &&
   -(n_2 + n_6)\,\delta(n_4)\,\Gamma(n_3 + n_5-1, n_2 + n_6-1, -2, w_2, w_3, 0)
\nonumber \\ &&
   +w_3\,\delta(n_4)\,\Gamma(n_3 + n_5-1, n_2 + n_6, -2, w_2, w_3, 0)
\nonumber \\ &&
   +n_6\,\delta(n_3)\,\Gamma(n_4 + n_5-1, n_6-1, n_2-2, w_1 + w_2, w_3, 0)
\nonumber \\ &&
   -w_3\,\delta(n_3)\,\Gamma(n_4 + n_5-1, n_6, n_2-2, w_1 + w_2, w_3, 0)
\nonumber \\ &&
   +n_2\,\delta(n_5)\,\Gamma(n_6-2, n_3 + n_4-1, n_2-1, w_3, w_1, 0)\Bigr],
\label{46}
\end{eqnarray}
where $n_i \geq 0$, and the formula for $f(-2, n_2, 1 + n_3, n_4, n_5, n_6)$ can be obtained from
the above one using symmetries of the function $f$. These recursions assumes that values of
$f(-2,0,0;n_4,n_5,n_6)$ are known. We again obtain them using IBP identities.
These are 9 equations, which we solve against the following $X_{i=1,9}$
unknowns at $u_2=u_3=0$
\begin{eqnarray}
X_1 &=&G(1,2,1,1,1,1)\,u_1^2\,, \nonumber\\
X_2 &=&G(1,1,2,1,1,1)\,u_1^2\,, \nonumber\\
X_3 &=&G(1,1,1,1,2,1)\,u_1^2\,, \nonumber\\
X_4 &=&G(1,2,1,1,1,2)\,u_1^2\,, \nonumber\\
X_5 &=&G(1,2,1,1,1,1)\,, \nonumber\\
X_6 &=&G(1,1,2,1,1,1)\,, \nonumber\\
X_7 &=&G(1,1,1,2,1,1)\,, \nonumber\\
X_8 &=&G(1,1,1,1,2,1)\,, \nonumber\\
X_9 &=&G(1,1,1,1,1,2). \label{47}
\end{eqnarray}
Equations for $X_7,\,X_8,$ and $X_9$ are
\begin{eqnarray}
0 &=&  (3\,w_1^2 + w_2^2 - w_3^2)\,G(1, 1, 1, 1, 1, 1) - 2\,u_1^2\,w_1^2\,G(2, 1, 1, 1, 1, 1)
\nonumber \\ &&
      - 4\,w_1^2\,w_2^2\,G(1, 1, 1, 1, 2, 1) - 2\,w_1^2\,(w_1^2 + w_2^2 - w_3^2)\,G(1, 1, 1, 2, 1, 1)
\nonumber \\ &&
 + F_1(u_1)\,w_2 - F_2(u_1)\,w_3\,, \label{48}\\
0 &=&   (3\,w_1^2 - w_2^2 + w_3^2)\,G(1, 1, 1, 1, 1, 1) - 2\,u_1^2\,w_1^2\,G(2, 1, 1, 1, 1, 1)
\nonumber \\ &&
       - 4\,w_1^2\,w_3^2\,G(1, 1, 1, 1, 1, 2) - 2\,w_1^2\,(w_1^2 - w_2^2 + w_3^2)\,G(1, 1, 1, 2, 1, 1)
\nonumber \\ &&
       -F_1(u_1)\,w_2 + F_2(u_1)\,w_3\,, \label{49}\\
0 &=& (3\,w_1^4 - 4\,w_1^2\,w_2^2 + w_2^4 - 4\,w_1^2\,w_3^2 - 2\,w_2^2\,w_3^2 + w_3^4)\,G(1, 1, 1, 1, 1, 1)
\nonumber \\ &&
     - 2\,w_1^2\,(w_1^4 - 2\,w_1^2\,w_2^2 + w_2^4 - 2\,w_1^2\,w_3^2 - 2\,w_2^2\,w_3^2
      + w_3^4)\,G(1, 1, 1, 2, 1, 1)
\nonumber \\ &&
      - 2\,w_1^2\,(u_1^2\,(w_1^2 - \,w_2^2 - \,w_3^2) + 2\,w_2^2\,w_3^2)\,G(2, 1, 1, 1, 1, 1)
\nonumber \\ &&
      - F_1(u_1)\,w_2\,(w_1^2 - w_2^2 + w_3^2) - F_2(u_1)\,w_3\,(w_1^2 + w_2^2 - w_3^2)\,,
\label{50}
\end{eqnarray}
where
\begin{eqnarray}
F_1(u_1) &=& 2\,w_2\,\bigl[G(1, 1, 1, 0, 2, 1) - G(1, 1, 1, 1, 2, 0) + G(2, 0, 1,
1, 1, 1) - G(2, 1, 1, 1, 1, 0)\bigr]\,,\nonumber \\
\label{51} \\
F_2(u_1) &=& 2\,w_3\,\bigl[G(1, 1, 1, 0, 1, 2) - G(1, 1, 1, 1, 0, 2) + G(2, 1, 0, 1, 1, 1) - G(2, 1, 1, 1, 0, 1)\bigr]\,.
\nonumber \\\label{52}
\end{eqnarray}
One performs the $u_1$ integration and obtains
\begin{eqnarray}
0 &=& F_1\,w_2 - F_2\,w_3 + (w_1^2 + w_2^2 - w_3^2)\,f(-2, 0, 0, 0, 0, 0)
\nonumber \\ &&
- 2\,w_1^2\,w_2\,f(-2, 0, 0, 0, 1, 0) - w_1\,(w_1^2 + w_2^2 - w_3^2)\,f(-2, 0, 0, 1, 0, 0)
\nonumber \\ &&
   + w_1^2\,\Gamma(0, -1, -1, w_1, w_2 + w_3, 0)\,, \label{53}\\
 0 &=& -F_1\,w_2 + F_2\,w_3 + (w_1^2 - w_2^2 + w_3^2)\,f(-2, 0, 0, 0, 0, 0)
\nonumber \\ &&
   - 2\,w_1^2\,w_3\,f(-2, 0, 0, 0, 0, 1) - w_1\,(w_1^2 - w_2^2 + w_3^2)\,f(-2, 0, 0, 1, 0, 0)
\nonumber \\ &&
    + w_1^2\,\Gamma(0, -1, -1, w_1, w_2 + w_3, 0)\,, \label{54} \\
0 &=& -F_2\,w_3\,(w_1^2 + w_2^2 - w_3^2) - F_1\,w_2\,(w_1^2 - w_2^2 + w_3^2) +
  w\,f(-2, 0, 0, 0, 0, 0)
\nonumber \\ &&
- w\,w_1\,f(-2, 0, 0, 1, 0, 0) - 2\,w_1^2\,w_2^2\,w_3^2\,f(0, 0, 0, 0, 0, 0)
\nonumber \\ &&
+ w_1^2\,(w_1^2 - w_2^2 - w_3^2)\,\Gamma(0, -1, -1, w_1, w_2 + w_3, 0)\,,\label{55}
\end{eqnarray}
where $w$ is defined in Eq. (\ref{18}) and
\begin{equation}
F_i = \int_0^\infty du_1\,u_1\,F_i(u_1)\label{56}
\end{equation}
in the sense of the integral defined in Eq. (\ref{30}), with the result
\begin{eqnarray}
F_1 &=&\frac{1}{2}\,\biggl[\ln\biggl(1 + \frac{w_2}{w_3}\biggr)\,\ln\biggl(\frac{w_2}{w_3}\biggr)
- \ln^2\biggl(\frac{w_3}{w_1}\biggr) + 2\,\ln\biggl(\frac{w_1}{w_2 + w_3}\biggr)
- 2\,\ln(w_3)\,\ln\biggl(\frac{w_1}{w_2 + w_3}\biggr)
\nonumber \\ &&
- \ln\biggl(\frac{w_2}{w_1 + w_3}\biggr)\,\ln\biggl(1 + \frac{w_2}{w_1 + w_3}\biggr) -
   2\,{\rm Li}_2\biggl( -\frac{w_2}{w_1}\biggr)
   + {\rm Li}_2\biggl( 1 - \frac{w_2}{w_3}\biggr)
   - {\rm Li}_2\biggl( -\frac{w_2}{w_3}\biggr)
\nonumber \\ &&
   - {\rm Li}_2\biggl( -\frac{w_2}{w_1 + w_3}\biggr)
   - {\rm Li}_2\biggl( 1 - \frac{w_2}{w_1 + w_3}\biggr)\biggr]\label{57}\\
    &=& \Gamma(0,-1,-2,w_2,w_3,0)  - \Gamma(0,-1,-2,0,w_1,w_2)
\nonumber \\ &&
        - w_2\,\Gamma(0,-1,-1,0,w_1,w_2) + w_2\,\Gamma(0,-1,-1,w_2,w_1+w_3,0)\,,\label{58}\\
        \nonumber \\
F_2 &=&\frac{1}{2}\,\biggl[\ln\biggl(1 + \frac{w_3}{w_2}\biggr)\,\ln\biggl(\frac{w_3}{w_2}\biggr)
- \ln^2\biggl(\frac{w_2}{w_1}\biggr) + 2\,\ln\biggl(\frac{w_1}{w_2 + w_3}\biggr)
- 2\,\ln(w_2)\,\ln\biggl(\frac{w_1}{w_2 + w_3}\biggr)
\nonumber \\ &&
- \ln\biggl(\frac{w_3}{w_1 + w_2}\biggr)\,\ln\biggl(1 + \frac{w_3}{w_1 + w_2}\biggr) -
   2\,{\rm Li}_2\biggl( -\frac{w_3}{w_1}\biggr)
   + {\rm Li}_2\biggl( 1 - \frac{w_3}{w_2}\biggr)
   - {\rm Li}_2\biggl( -\frac{w_3}{w_2}\biggr)
\nonumber \\ &&
   - {\rm Li}_2\biggl( -\frac{w_3}{w_1 + w_2}\biggr)
   - {\rm Li}_2\biggl( 1 - \frac{w_3}{w_1 + w_2}\biggr)\biggr]\label{59}\\
   &=&\Gamma(0,-1,-2,w_3,w_2,0) - \Gamma(0,-1,-2,0,w_1,w_3)
\nonumber \\ &&
   - w_3\,\Gamma(0,-1,-1,0,w_1,w_3) + w_3\,\Gamma(0,-1,-1,w_3,w_1+w_2,0)\,.\label{60}
\end{eqnarray}
Under the replacement $f(-2, 0, 0, 0, 0, 0) = h/w_1\equiv h(0,0,0)/w_1$
one obtains simplified differential equations
\begin{eqnarray}
0 &=&  2\,w_1\,w_2\,\frac{\partial h}{\partial w_2}
+ (w_1^2 + w_2^2 - w_3^2)\,\frac{\partial h}{\partial w_1}
+ F_1\,w_2 - F_2\,w_3
+ w_1^2\,\Gamma(0, -1, -1, w_1, w_2 + w_3, 0)\,,\nonumber \\ \label{61}\\
 0 &=& 2\,w_1\,w_3\,\frac{\partial h}{\partial w_3}
+ (w_1^2 - w_2^2 + w_3^2)\,\frac{\partial h}{\partial w_1}
-F_1\,w_2 + F_2\,w_3
+ w_1^2\,\Gamma(0, -1, -1, w_1, w_2 + w_3, 0)\,, \nonumber \\ \label{62}\\
0 &=& w\,\frac{\partial h}{\partial w_1}
-F_2\,w_3\,(w_1^2 + w_2^2 - w_3^2) - F_1\,w_2\,(w_1^2 - w_2^2 + w_3^2)
 - 2\,w_1^2\,w_2^2\,w_3^2\,f(0, 0, 0, 0, 0, 0)
\nonumber \\ &&
+ w_1^2\,(w_1^2 - w_2^2 - w_3^2)\,\Gamma(0, -1, -1, w_1, w_2 + w_3, 0)\,.\label{63}
\end{eqnarray}
One differentiates Eq. (\ref{61}) with respect to $w_1$, $w_2$, and $w_3$,
and obtains in this way recursions for $h$ in $n_5$
\begin{eqnarray}
h(n_4, n_5+1, n_6) &=& \frac{1}{2\,w_1\,w_2}\,\Bigl[
-n_5\,F_1(n_4, n_5-1, n_6) + w_2\,F_1(n_4, n_5, n_6)
\nonumber \\ &&
+ n_6\,F_2(n_4, n_5, n_6-1) - w_3\,F_2(n_4, n_5, n_6)
\nonumber \\ &&
+ (n_4-1)\,n_4\,\Gamma(-1, n_4-2, n_5 + n_6-1, 0, w_1, w_2 + w_3)
\nonumber \\ &&
- 2\,n_4\,w_1\,\Gamma(-1, n_4-1, n_5 + n_6-1, 0, w_1, w_2 + w_3)
\nonumber \\ &&
+ w_1^2\,\Gamma(-1, n_4, n_5 + n_6-1, 0, w_1, w_2 + w_3)
\nonumber \\ &&
- n_4\,(n_4 + 2\,n_5-1)\,h(n_4-1, n_5, n_6)
\nonumber \\ &&
+ 2\,n_4\,w_2\,h(n_4-1, n_5+1, n_6)
\nonumber \\ &&
+ 2\,(n_4 + n_5)\,w_1\,h(n_4, n_5, n_6)
\nonumber \\ &&
- (n_5-1)\,n_5\,h(n_4+1, n_5-2, n_6)
\nonumber \\ &&
+ 2\,n_5\,w_2\,h(n_4+1, n_5-1, n_6)
\nonumber \\ &&
+ (n_6-1)\,n_6\,h(n_4+1, n_5, n_6-2)
\nonumber \\ &&
- 2\,n_6\,w_3\,h(n_4+1, n_5, n_6-1)
\nonumber \\ &&
- (w_1^2 + w_2^2 - w_3^2)\,h(n_4+1, n_5, n_6)\Bigr]\,,\label{64}
\end{eqnarray}
where
\begin{eqnarray}
F_1(n_4,n_5,n_6) &=&
\delta(n_4)\,\Gamma(-2, n_5, n_6-1, 0, w_2, w_3)
\nonumber \\ &&
- n_5\,\Gamma(-1, n_5-1, n_4 + n_6-1, 0, w_2, w_1 + w_3)
\nonumber \\ &&
+ w_2\,\Gamma(-1, n_5, n_4 + n_6-1, 0, w_2, w_1 + w_3)
\nonumber \\ &&
+ (n_5-1)\,\delta(n_6)\,\Gamma(0, n_4-1, n_5-2, 0, w_1, w_2)\,,
\nonumber \\ \label{65}&&
- w_2\,\delta(n_6)\,\Gamma(0, n_4-1, n_5-1, 0, w_1, w_2),\\
F_2(n_4,n_5,n_6) &=&
 \delta(n_4)\,\Gamma(-2, n_5-1, n_6, 0, w_2, w_3)
\nonumber \\ &&
- n_6\,\Gamma(-1, n_6-1, n_4 + n_5-1, 0, w_3, w_1 + w_2)
\nonumber \\ &&
+ w_3\,\Gamma(-1, n_6, n_4 + n_5-1, 0, w_3, w_1 + w_2)
\nonumber \\ &&
+ (n_6-1)\,\delta(n_5)\,\Gamma(0, n_4-1, n_6-2, 0, w_1, w_3)
\nonumber \\ &&
- w_3\,\delta(n_5)\,\Gamma(0, n_4-1, n_6-1, 0, w_1, w_3)\,,\label{66}
\end{eqnarray}
and the formula for $h(n_4, n_5, 1 + n_6)$ is analogous.
Eq. (\ref{63}) is transformed now to the more explicit form
\begin{eqnarray}
\frac{\partial h}{\partial w_1} &=& \frac{g_1}{w_1+w_2+w_3} +
\frac{g_2}{w_1-w_2-w_3}+\frac{g_3}{w_1-w_2+w_3}+\frac{g_4}{w_1+w_2-w_3}\,,\label{67}\\ \nonumber \\
g_1 &=&\frac{1}{4}\,\biggl[-\frac{\pi^2}{6} +
\ln\biggl(\frac{w_1}{w_2}\biggr)\,\ln\biggl(\frac{w_1}{w_3}\biggr) +
\bigl[\ln(w_2\,w_3)-2\bigr]\,\ln\biggl(\frac{w_1}{w_2+w_3}\biggr)
\nonumber \\ &&
+ {\rm Li}_2\biggl( -\frac{w_2}{w_1}\biggr) + {\rm Li}_2\biggl(-\frac{w_3}{w_1}\biggr)\biggr]\,,
\label{68}\\ \nonumber \\
g_2 &=&\frac{1}{4}\,\biggl[-\ln\biggl(\frac{w_1}{w_2}\biggr)\,\ln\biggl(\frac{w_1}{w_3}\biggr) +
\bigl[2-\ln(w_2\,w_3)\bigr]\,\ln\biggl(\frac{w_1}{w_2+w_3}\biggr) \nonumber \\ &&
- {\rm Li}_2\biggl( -\frac{w_2}{w_1}\biggr)
- {\rm Li}_2\biggl( -\frac{w_3}{w_1}\biggr)
-{\rm Lg}\biggl(\frac{w_3}{w_1 + w_2}\biggr)
- {\rm Lg}\biggl(\frac{w_2}{w_1 + w_3}\biggr)\biggr]\,,
\label{69} \\\nonumber  \\
g_3 &=&\frac{1}{8}\,\biggl[-\frac{\pi^2}{6}
+ 2\,{\rm Lg}\biggl(\frac{w_3}{w_1 + w_2}\biggr)
+ 2\,{\rm Lg}\biggl(\frac{w_1}{w_2 + w_3}\biggr)
- 2\,{\rm Li}_2\biggl( -\frac{w_2}{w_1}\biggr)
+ {\rm Li}_2\biggl( 1 - \frac{w_2}{w_3}\biggr)
\nonumber \\ &&
- {\rm Li}_2\biggl( -\frac{w_2}{w_3}\biggr)
+ 2\,{\rm Li}_2\biggl( -\frac{w_3}{w_1}\biggr)
+ {\rm Li}_2\biggl( -\frac{w_3}{w_2}\biggr)
- {\rm Li}_2\biggl( 1 - \frac{w_3}{w_2}\biggr)\biggr]\,, \label{70}\\ \nonumber \\
g_4 &=&\frac{1}{8}\,\biggl[-\frac{\pi^2}{6}
+ 2\,{\rm Lg}\biggl(\frac{w_2}{w_1 + w_3}\biggr)
+ 2\,{\rm Lg}\biggl(\frac{w_1}{w_2 + w_3}\biggr)
+ 2\,{\rm Li}_2\biggl( -\frac{w_2}{w_1}\biggr)
- {\rm Li}_2\biggl( 1 - \frac{w_2}{w_3}\biggr)
\nonumber \\ &&
+ {\rm Li}_2\biggl( -\frac{w_2}{w_3}\biggr)
- 2\,{\rm Li}_2\biggl( -\frac{w_3}{w_1}\biggr)
- {\rm Li}_2\biggl( -\frac{w_3}{w_2}\biggr)
+ {\rm Li}_2\biggl( 1 - \frac{w_3}{w_2}\biggr)\biggr]\,, \label{71}
\end{eqnarray}
These formulae are used to obtain forward recursions for $h(n_4,0,0)$
similar to those used in two-electron Hylleraas integrals \cite{gamma}.
Namely, if we denote by $h_i = g_i/(w_1 +\ldots)$ the corresponding term in
Eq. (\ref{67}), then the recursion for example for $h_2(n)$ defined by
\begin{equation}
h_2(n) = \biggl(-\frac{\partial}{\partial w_1}\biggr)^n h_2\,,
\end{equation}
in the general case is the following
\begin{equation}
h_2(n) = \frac{g_2(n)+n\,h_2(n-1)}{w_1-w_2-w_3}\,.
\end{equation}
In the special case of $w_1\approx w_2+w_3$
one calculates at first recursions exactly at $w_1= w_2+w_3$
\begin{equation}
h_2(n) = -\frac{g_2(n+1)}{n+1}\,,
\end{equation}
and obtains $h_2(n)$ for $w_1\neq w_2+w_3$ using the Taylor formula.
For this one notes that first derivatives of $g_i$ are quite simple,
\begin{eqnarray}
\frac{\partial g_1}{\partial w_1} &=& \frac{1}{4\,w_1}\,\bigl[\ln(w_1+w_2)+\ln(w_1+w_3)-2\bigr],\label{72}\\
\frac{\partial g_2}{\partial w_1} &=&-\frac{1}{4\,w_1}\,\bigl[\ln(w_1+w_2)+\ln(w_1+w_3)-2\bigr]
+\frac{1}{4\,(w_1+w_2-w_3)}\,\ln\biggl(\frac{w_3}{w_1+w_2}\biggr)+
\label{73}\\ &&
\frac{1}{4\,(w_1-w_2+w_3)}\,\ln\biggl(\frac{w_2}{w_1+w_3}\biggr)
-\frac{1}{4\,(w_1+w_2+w_3)}\,\biggl[\ln\biggl(\frac{w_3}{w_1+w_2}\biggr)
+\ln\biggl(\frac{w_2}{w_1+w_3}\biggr)\biggr]\,,
\nonumber\\
\frac{\partial g_3}{\partial w_1} &=&
\frac{1}{4\,w_1}\,\ln\biggl(\frac{w_1+w_3}{w_1+w_2}\biggr)
-\frac{1}{4\,(w_1+w_2-w_3)}\,\ln\biggl(\frac{w_3}{w_1+w_2}\biggr)-
\label{74}\\ &&
\frac{1}{4\,(w_1-w_2-w_3)}\,\ln\biggl(\frac{w_1}{w_2+w_3}\biggr)
+\frac{1}{4\,(w_1+w_2+w_3)}\,\biggl[\ln\biggl(\frac{w_3}{w_1+w_2}\biggr)
+\ln\biggl(\frac{w_1}{w_2+w_3}\biggr)\biggr]\,,
\nonumber\\
\frac{\partial g_4}{\partial w_1} &=& \frac{1}{4\,w_1}\,\ln\biggl(\frac{w_1+w_2}{w_1+w_3}\biggr)
-\frac{1}{4\,(w_1-w_2+w_3)}\,\ln\biggl(\frac{w_2}{w_1+w_3}\biggr)-
\label{75}\\ &&
\frac{1}{4\,(w_1-w_2-w_3)}\,\ln\biggl(\frac{w_1}{w_2+w_3}\biggr)
+\frac{1}{4\,(w_1+w_2+w_3)}\,\biggl[\ln\biggl(\frac{w_2}{w_1+w_3}\biggr)
+\ln\biggl(\frac{w_1}{w_2+w_3}\biggr)\biggr]\,.
\nonumber
\end{eqnarray}
Independly of recursions one needs the function $h$.
It is obtained from Eq. (\ref{67})
\begin{equation}
h = \int_0^{w_1}\,dw_1\,\frac{\partial h}{\partial w_1},\label{76}
\end{equation}
and this integral can be explicitly performed in terms of Li$_3$, Li$_2$
and logarithmic functions, but the result is quite complicated. For
numerical calculations, the integral representation of Eq. (\ref{76})
with the adapted quadrature \cite{rokhlin} is used to achieve about 64 digits accuracy.

Having $h(n_4,n_5,n_6)$, the function $f(-2,0,0,n_4,n_5,n_6)$ is obtained by
\begin{equation}
f(-2,0,0,n_4,n_5,n_6) = \frac{1}{w_1}\bigl[h(n_4,n_5,n_6)+n_4\,f(-2,0,0,n_4-1,n_5,n_6)\bigr],\label{77}
\end{equation}
which completes the evaluation of singular Hylleraas integrals.

\section{Summary}
We have obtained recursion relations for three-electron Hylleraas integrals
involving $1/r_i^3$ and $1/r_{ij}^3$. The initial multi dimensional master integral is
expressed in the form of a one dimensional integral, which is suitable for high
precision numerical evaluation. A set of numerical examples
for $f(-2,0,0,0,0,0)$ is presented in Table I and for $f(0,0,0,-2,0,0)$
in Table II. Together with regular and extended Hylleraas integrals
they allow for the calculation of leading QED effects in three electron atoms or
ions. Accurate results for the lithium atom have already been obtained by Yan and
Drake in \cite{lit2s3sdy} and several precise measuements
have been performed in \cite{lit2s3sexp,lit2s4sexp}. 
While we intend to verify their result, we aim also to
calculate QED effects in the Be$^+$ ion for the determination of the nuclear charge
radii from planned isotope shift measurements. Moreover, we aim
to calculate higher order QED effects for energy levels along with fine and hyperfine
splittings \cite{krp_lit} to verify certain discrepancies, for example the difference between
experimental values and theoretical predictions for the 2S-3S transition frequency
in Li \cite{lit2s3sexp, lit2s3sdy}.

\begin{table}[!hbt]
\caption{Values of the master integral $f(-2,0,0,0,0,0)$ for selected $w_1,w_2$, and $w_3$}
\label{table1}
\begin{ruledtabular}
\begin{tabular}{llll}
      $w_1$ & $w_2$  & $w_3$  &  $f(-2,0,0,0,0,0)$  \\
\hline
4.0 & 1.0 & 0.5 &   -0.712\,305\,106\,830\,898\,240\,034\,381\,269\,753\,396 \\
4.0 & 1.0 & 1.0 &   -0.699\,249\,073\,328\,162\,683\,321\,507\,663\,896\,763 \\
4.0 & 1.0 & 1.5 &   -0.691\,399\,920\,204\,079\,424\,318\,079\,206\,069\,445 \\
4.0 & 1.0 & 2.0 &   -0.680\,940\,098\,270\,060\,661\,183\,207\,296\,253\,486 \\
4.0 & 1.0 & 2.5 &   -0.667\,068\,654\,115\,517\,811\,166\,904\,215\,261\,918 \\
4.0 & 1.0 & 3.0 &   -0.651\,691\,691\,901\,107\,499\,331\,195\,660\,199\,574 \\
4.0 & 1.0 & 3.5 &   -0.640\,014\,560\,205\,730\,366\,159\,157\,581\,017\,121 \\
4.0 & 1.0 & 4.0 &   -0.627\,151\,433\,983\,103\,414\,178\,318\,921\,819\,093 \\
4.0 & 1.0 & 4.5 &   -0.614\,807\,615\,676\,494\,534\,607\,896\,770\,432\,226 \\
4.0 & 1.0 & 5.0 &   -0.602\,911\,457\,174\,426\,959\,108\,113\,476\,921\,161 \\
4.0 & 1.0 & 5.5 &   -0.591\,697\,977\,384\,241\,861\,061\,839\,402\,802\,568 \\
\\
\end{tabular}
\end{ruledtabular}
\end{table}

\begin{table}[!hbt]
\caption{Values of the master integral $f(0,0,0,-2,0,0)$ for selected $w_1,w_2$, and $w_3$}
\label{table2}
\begin{ruledtabular}
\begin{tabular}{llll}
      $w_1$ & $w_2$  & $w_3$  &  $f(0,0,0,-2,0,0)$  \\
\hline
4.0 & 1.0 & 0.5 &   -3.457\,452\,854\,159\,239\,112\,750\,191\,117\,485\,829 \\
4.0 & 1.0 & 1.0 &   -2.650\,037\,638\,635\,495\,670\,871\,341\,393\,412\,236 \\
4.0 & 1.0 & 1.5 &   -2.234\,238\,493\,311\,187\,793\,632\,537\,945\,604\,453 \\
4.0 & 1.0 & 2.0 &   -1.967\,448\,655\,334\,872\,044\,627\,131\,808\,094\,930 \\
4.0 & 1.0 & 2.5 &   -1.777\,079\,267\,715\,778\,085\,428\,017\,041\,995\,909 \\
4.0 & 1.0 & 3.0 &   -1.632\,246\,974\,419\,083\,835\,406\,405\,271\,425\,506 \\
4.0 & 1.0 & 3.5 &   -1.517\,186\,855\,826\,194\,722\,590\,295\,936\,030\,308 \\
4.0 & 1.0 & 4.0 &   -1.422\,872\,681\,980\,733\,011\,235\,327\,186\,357\,640 \\
4.0 & 1.0 & 4.5 &   -1.343\,706\,445\,824\,538\,744\,395\,243\,994\,507\,941 \\
4.0 & 1.0 & 5.0 &   -1.276\,004\,428\,705\,473\,405\,554\,500\,648\,664\,743 \\
4.0 & 1.0 & 5.5 &   -1.217\,228\,914\,605\,386\,522\,608\,906\,840\,145\,094 \\
\end{tabular}
\end{ruledtabular}
\end{table}

\section*{Acknowledgments}
The authors wish to acknowledge interesting discussions with Krzysztof Meissner
and Vladimir Korobov.
This work was supported by NIST precision measurement grant 60NANB7D6153.

\section*{Appendix: Singular two electron integrals }
The definition of the distribution $P(1/r^3)$ assumes
that it is integrated with a smooth function of $r$.
This is not always the case with Hylleraas integrals.
One can not apply directly the definition of Eq. (\ref{28}), because
the integrand is logarithmic in $r$  for certain negative powers
of electron coordinates.
Therefore, here we assume the following recursive definition
\begin{equation}
-\frac{\partial}{\partial\gamma}\,\Gamma(n_1,n_2,n_3,\alpha,\beta,\gamma) =
\Gamma(n_1,n_2,n_3+1,\alpha,\beta,\gamma) \label{78}
\end{equation}
and set the boundary condition for large $\gamma$,
by the requirement that only powers of $\ln\gamma$ are present, but no constant
term. Using this definition, which is consistent with Eq. (\ref{28}),
one obtains
\begin{eqnarray}
\Gamma(n_1,n_2,n_3,\alpha,\beta,\gamma)
&\equiv&\int\frac{d^3r_1}{4\,\pi}\,\int\frac{d^3r_2}{4\,\pi}\,
e^{-\alpha\,r_1-\beta\,r_2-\gamma\,r_{12}}\,r_1^{n_1-1}\,r_2^{n_2-1}\,r_{12}^{n_3-1}
\label{79}\\ \nonumber \\
\Gamma(0,0,0,\alpha,\beta,\gamma) &=& \frac{1}{(\alpha+\beta)\,(\alpha+\gamma)\,(\beta+\gamma)}
\label{80}\\ \nonumber \\
\Gamma(0,0,-1,\alpha,\beta,\gamma)  &=& \frac{1}{(\alpha-\beta)\,(\alpha+\beta)}
                                        \ln\biggl(\frac{\gamma+\alpha}{\gamma+\beta}\biggr)
\label{81}\\ \nonumber \\
\Gamma(0,0,-2,\alpha,\beta,\gamma)  &=&
\frac{(\beta+\gamma) \,\ln(\beta+\gamma)-(\alpha+\gamma)\,\ln(\alpha+\gamma)}{(\alpha^2-\beta^2)}+
                                        \frac{1}{\alpha + \beta}
\label{82}\\ \nonumber \\
\Gamma(0,-1,-1,0,\beta,\gamma)  &=& \frac{1}{\gamma}\,\ln\biggl(1+\frac{\gamma}{\beta}\biggr)
                                   +\frac{1}{\beta}\,\ln\biggl(1+\frac{\beta}{\gamma}\biggr)
\label{83}\\ \nonumber \\
\Gamma(0,-1,-1,\alpha,\beta,0) &=& \frac{1}{2\,\alpha}\,\biggl[\frac{\pi^2}{6}-
                                   \ln\biggl(1+\frac{\alpha}{\beta}\biggr)\,\ln\frac{\alpha}{\beta}-
                                   {\rm Li}_2\biggl(1-\frac{\alpha}{\beta}\biggr)-
                                   {\rm Li}_2\biggl(-\frac{\alpha}{\beta}\biggr)\biggr]
\label{84}\\ \nonumber \\
\Gamma(0,-1,-1,\alpha,\beta,\gamma) &=& \frac{1}{2\,\alpha}\,\biggl[\frac{\pi^2}{6}+
                                        \frac{1}{2}\,\ln^2\biggl(\frac{\alpha+\beta}{\alpha+\gamma}\biggr)
                                        +{\rm Li}_2\biggl(1-\frac{\beta+\gamma}{\alpha+\beta}\biggr)
                                        +{\rm Li}_2\biggl(1-\frac{\beta+\gamma}{\alpha+\gamma}\biggr)\biggr]
\nonumber \\ \label{85}\\
\Gamma(0,-1,-2,0,\beta,\gamma)  &=& {\rm  Li}_2\biggl(-\frac{\gamma}{\beta}\biggr)
                                    -\frac{\gamma}{\beta}\,\ln\biggl(1+\frac{\beta}{\gamma}\biggr)
                                    +\frac{1}{2}\,\ln^2\beta-\ln(\beta+\gamma)+\frac{\pi^2}{6}+1
\label{86}\\ \nonumber \\
\Gamma(0,-1,-2,\alpha,\beta,0)  &=& \frac{1}{2}\,{\rm  Li}_2\biggl(1-\frac{\alpha}{\beta}\biggr)
                                   -\frac{1}{2}\,{\rm  Li}_2\biggl(-\frac{\alpha}{\beta}\biggr)
                                   +\frac{1}{2}\,\ln\alpha\,\ln\biggl(1+\frac{\alpha}{\beta}\biggr)
\nonumber \\ &&
                                   +\biggl(\frac{\ln\beta}{2}-1\biggr)\,\ln(\alpha+\beta)
                                   +\frac{\pi^2}{12}+1
\label{87}\\ \nonumber \\
\Gamma(0,-1,-2,\alpha,\beta,\gamma) &=&
\frac{1}{2}\,\biggl(1-\frac{\gamma}{\alpha}\biggr)\,{\rm Li}_2\biggl(1-\frac{\beta+\gamma}{\alpha+\beta}\biggr)
-\frac{1}{2}\,\biggl(1+\frac{\gamma}{\alpha}\biggr)\,{\rm Li}_2\biggl(1-\frac{\beta+\gamma}{\alpha+\gamma}\biggr)
\nonumber \\ &&
-\frac{1}{4}\,\biggl(1+\frac{\gamma}{\alpha}\biggr)\,\ln^2\biggl(\frac{\alpha+\gamma}{\alpha+\beta}\biggr)
+\frac{1}{2}\,\ln^2(\alpha+\beta)-\ln(\alpha+\beta)
\nonumber \\ &&
+\frac{\pi^2}{12}\,\biggl(1-\frac{\gamma}{\alpha}\biggr) +1 \label{88}
\end{eqnarray}


\begin{thebibliography}{99}

\bibitem{lit_wave1} Z.-C. Yan and G. W. F. Drake, Phys. Rev. A {\bf 52}, 3711 (1995).

\bibitem{lit_wave2} M. Puchalski and K. Pachucki, Phys. Rev. A {\bf 73}, 022503 (2006).

\bibitem{hyll} E.A.~Hylleraas, Z. Phys. {\bf 54}, 347 (1929).

\bibitem{king_lit_nrel}
F. W. King and V. Shoup, Phys. Rev. A {\bf 33}, 2940 (1986); F. W.
King and M. P. Bergsbaken, J. Chem. Phys. {\bf 93}, 2570 (1990);
F. W. King, J. Chem. Phys. {\bf 99}, 3622 (1993); F. W. King,
THEOCHEM: J. Mol. Struct. {\bf 400}, 7 (1997); P.J. Pelzl and F.W.
King, Phys. Rev. E {\bf 57}, 7268 (1998); F.W. King, Adv. At. Mol.
Opt. Phys. {\bf 40}, 57 (1999).

\bibitem{yandrake_lit_nrel}
        D. K. McKenzie and G. W. F. Drake, Phys. Rev. A {\bf 44}, R6973 (1991);
        G.W.F. Drake and Z.-C. Yan, Phys. Rev. A {\bf 46}, 2378 (1992).

\bibitem{sims} J. S. Sims and S. A. Hagstrom, J. Phys. B: At. Mol. Opt. Phys. 
               {\bf 37}, 1519 (2004); {\em ibid}. {\bf 40}, 1575 (2007).

\bibitem{recursions}K. Pachucki, M. Puchalski and E. Remiddi,
                    Phys. Rev. A {\bf 70}, 032502 (2004).

\bibitem{king_lit_sing}
 D.M. Feldman, P.J. Pelzl, and F. W. King, J. Math. Phys. {\bf 39}, 6262 (1998);
 F. W. King {\it et al.} Phys. Rev. A {\bf 58}, 3597 (1998);
        F. W. King, Int. J. Quant. Chem. {\bf 72}, 93 (1999);
        P.J. Pelzl, G.J. Smethells, and F.W. King, Phys. Rev. E {\bf 65}, 036707 (2002).

\bibitem{yandrake_lit_sing}
        G. W. F. Drake and Z.-C. Yan, Phys. Rev. A {\bf 52}, 3681 (1995);
        J. Phys. B: At. Mol. Opt. Phys. {\bf 30}, 4723 (1997);
        Z.-C. Yan and G. W. F. Drake, Phys. Rev. A {\bf 61}, 022504 (2000);
        Phys. Rev. A {\bf 66}, 042504 (2002);
        Phys. Rev. Lett. {\bf 91}, 113004 (2003).

\bibitem{rec_sing} K. Pachucki and M. Puchalski, Phys. Rev. A {\bf 71}, 032514 (2005),

\bibitem{yan} Z.-C. Yan, J. Phys. B: At. Mol. Opt. Phys. {\bf 33}, 2437 (2000). 

\bibitem{remiddi} E.~Remiddi, \PRA{44}{5492}{1991}

\bibitem{rokhlin} J. Ma, V. Rokhlin, and S. Wandzura, SIAM
  J. Numer. Anal. {\bf 33}, 971 (1996).

\bibitem{parts}     F.V. Tkachov, Phys. Lett. B {\bf 100}, 65 (1981);
                    K.G. Chetyrkin and F.V. Tkachov, Nucl. Phys. B {\bf 192},
                    159 (1981).

\bibitem{gamma} R.A. Sack, C.C.J Roothaan and W. Ko\l os, J. Math. Phys. {\bf 8}, 1093 (1967);
                V.I. Korobov, J. Phys. B {\bf 35}, 1959 (2002);
                F.E Harris, A.M. Frolov and V.S. Smith, Jr., J. Chem. Phys {\bf 121}, 6323 (2004).

\bibitem{lit2s4sexp} W. DeGraffenreid and C.J. Sansonetti, Phys. Rev. A {\bf 67}, 012509 (2003).

\bibitem{lit2s3sexp} B. A. Bushaw, W. N\"ortersh\"auser, G. Ewald, A. Dax, and
  G. W. F. Drake, Phys. Rev. Lett. {\bf 91}, 043004 (2003).

\bibitem{krp_lit} K. Pachucki, Phys. Rev. A {\bf 66}, 062501 (2002).

\bibitem{lit2s3sdy} Z.-C. Yan and G. W. F. Drake, Phys. Rev. A 66, 042504 (2002),
  Phys. Rev. Lett. {\bf 91}, 113004 (2003).

\end{thebibliography}
\end{document}